\documentclass{aa}

\usepackage{graphicx}
\usepackage{times}
\usepackage{float}

\begin{document}

   \thesaurus{06         
              (13.07.1)}  
            \title{The extraordinarily bright optical afterglow of GRB 991208 and its host galaxy}



   \author{A. J. Castro-Tirado
          \inst{1,2}
   \and V. V. Sokolov
          \inst{3,4}
   \and J. Gorosabel
          \inst{5}
   \and J. M. Castro Cer\'on
          \inst{6}
   \and J. Greiner
          \inst{7}
   \and R. A. M. J. Wijers
          \inst{8}
   \and B. L. Jensen
          \inst{9}
   \and J. Hjorth
          \inst{9}
   \and S. Toft
          \inst{9}
   \and H. Pedersen
          \inst{9}
   \and E. Palazzi
          \inst{10}
   \and E. Pian
          \inst{10}
   \and N. Masetti
          \inst{10}
   \and R. Sagar
          \inst{11}
   \and V. Mohan
          \inst{11}
   \and A. K. Pandey
          \inst{11}
   \and S. B. Pandey
          \inst{11}
   \and S. N. Dodonov
          \inst{3}
   \and T. A. Fatkhullin
          \inst{3}
   \and V. L. Afanasiev
          \inst{3}
   \and V. N. Komarova
          \inst{3,4}
   \and A. V. Moiseev
          \inst{3}
   \and R. Hudec
          \inst{12}
   \and V. Simon
          \inst{12}
   \and P. Vreeswijk
          \inst{13}
   \and E. Rol
          \inst{13}
   \and S. Klose
          \inst{14}
   \and B. Stecklum
          \inst{14}
   \and M. R. Zapatero-Osorio
          \inst{15}
   \and N. Caon
          \inst{15}
   \and C. Blake
          \inst{16}
   \and J. Wall
          \inst{16}
   \and D. Heinlein
          \inst{17}
   \and A. Henden
          \inst{18,19}
   \and S. Benetti
          \inst{20}
   \and A. Magazz\`u
          \inst{20}
   \and F. Ghinassi
          \inst{20}
   \and L. Tommasi
          \inst{21}
   \and M. Bremer
          \inst{22}
   \and C. Kouveliotou
          \inst{23}
   \and S. Guziy
          \inst{24}
   \and A. Shlyapnikov
          \inst{24}
   \and U. Hopp
          \inst{25}
   \and G. Feulner
          \inst{25}
   \and S. Dreizler
          \inst{26}
   \and D. Hartmann
          \inst{27}
   \and H. Boehnhardt
          \inst{28}
   \and J. M. Paredes 
          \inst{29}
   \and J. Mart\'{\i}
          \inst{30}
   \and E. Xanthopoulos
          \inst{31}
   \and H. E. Kristen
          \inst{32}
   \and J. Smoker
          \inst{33}
   \and K. Hurley
          \inst{34}
          }

   \offprints{A. J. Castro-Tirado (ajct@laeff.esa.es)}

   \institute{Instituto de Astrof\'{\i}sica de Andaluc\'{\i}a (IAA-CSIC), 
              P.O. Box 03004, E-18080 Granada, Spain
   \and       Laboratorio de Astrof\'{\i}sica Espacial y F\'{\i}sica 
              Fundamental (LAEFF-INTA), 
              P.O. Box 50727, E-28080 Madrid, Spain   
   \and Special Astrophysical Observatory of the Russian Academy of Sciences, Karanchai-Cherkessia, Nizhnij Arkhyz, 357147, Russia
   \and Isaac Newton Institute of Chile, SAO Branch
   \and Danish Space Research Institute, Copenhagen, Denmark
   \and Real Instituto y Observatorio de la Armada, Secci\'on de 
Astronom\'{\i}a, 11110 San Fernando--Naval, C\'adiz, Spain
   \and Astrophysikalisches Institut, Potsdam, Germany
   \and Department of Physics and Astronomy, SUNY Stony Brook, NY, USA
   \and Astronomical Observatory, University of Copenhagen, Copenhagen, 
Denmark
   \and Istituto Tecnologie e Studio Radiazioni Extraterrestri, 
CNR, Bologna, Italy
   \and U. P. State Observatory, Manora Peak, Nainital, 263 129 India
   \and Astronomical Institute of the Czech Academy of Sciences,
              251 65 Ondrejov, Czech Republic
   \and Anton Pannekoehk Institut, Amsterdam, Netherlands
   \and Th\"uringer Landessternwarte, Sternwarte 5, D-07778 Tautenburg, 
Germany
   \and Instituto de Astrof\'{\i}sica de Canarias, La Laguna, Tenerife, 
Spain
   \and Oxford University, AX1 4AU Oxford, United Kingdom
   \and Deutsches Zentrum f\"ur Luft- und Raumfahrt, Lilienstrasse 3. D 86156 Augsburg, Germany
   \and U. S. Naval Observatory, Flagstaff station, AZ, USA
   \and Universities Space Research Association, Flagstaff station, AZ, U.S.A.
   \and Centro Galileo Galilei, Canary Islands, Spain
   \and Universit\'a di Milano, Dipartimento di Fisica, Via Celoria 16, I-20133 Milano, Italia
   \and Institut de Radio Astronomie Millimetrique, Grenoble, France
   \and Universities Research Association, SD-50, NASA/MSFC, Hunstville, AL 
35812, USA
   \and Nikolaev University Observatory, Nikolskaya 24, 327030 Nikolaev, 
    Ukraine
   \and Universit\"ats-Sternwarte,  M\"unchen,  Germany
   \and University of T\"ubingen, T\"ubingen, Germany
   \and Clemson University, Department of Physics and Astronomy, 
Clemson, SC 29634, USA
   \and European Southern Observatory, Santiago, Chile
   \and Departmeent d\' \rm Astronimia i Meteorologia, Universidad de Barcelona, Avda. Diagonal 647, E-08028 Barcelona, Spain
   \and Departamento de F\'{\i}sica, Escuela Polit\'ecnica Superior, Universidad de Ja\'en, Virgen de la Cabeza 2, E-23071 Ja\'en, Spain
   \and University of Manchester, Jodrell Bank Observatory, Macclesfield, Cheshire, SK11 9DL, England
   \and Harvard - Smithsonian Center for Astrophysics, Harvard, USA
   \and Department of Pure and Applied Physics, Queens University Belfast, University Road, Belfast, BT7 1NN, United Kingdom
   \and Space Science Laboratory, University of California at Berkerley, USA  
             }

\date{Received date; accepted date}
\thesaurus{03(11.04.1,11.07.1,13.07.1)}


\maketitle
\markboth{A. J. Castro-Tirado et al.:The extraordinarily bright optical 
afterglow of GRB 991208.}{}

\begin{abstract}

Broad-band optical observations of the extraordinarily bright optical 
afterglow of the intense gamma-ray burst GRB 991208 started $\sim$ 2.1 
days after the event and continued until 4 Apr 2000.
The flux decay constant of the optical afterglow in the R-band is
$-$2.30 $\pm$ 0.07 up to $\sim$ 5 days, which is very likely due to the jet 
e\-ffect, and after that it is followed by a much steeper decay 
with constant $-$3.2 $\pm$ 0.2, the fastest one ever seen in a GRB
optical afterglow. A 
negative detection in several all-sky films taken simultaneously to 
the event, that otherwise would have reached naked eye brightness, implies
either a previous additional break prior to $\sim$ 2 days after the 
occurrence of the GRB (as expected from the jet effect) or a maximum as 
observed in GRB 970508. The existence of a se\-cond break might indicate a 
steepening in the electron spectrum or the superposition of two events, 
resembling GRB 000301C. Once the afterglow emission vanished, 
contribution of a bright underlying supernova is found on the basis of 
the late-time R-band measurements, but the light curve is not sufficiently
well sampled to rule out a dust echo explanation.
Our redshift determination of z = 0.706 indicates that GRB 991208 is at 
3.7 Gpc (for H$_{0}$ = 60 km s$^{-1}$ Mpc$^{-1}$, $\Omega_{0}$ = 1 and 
$\Lambda_{0}$ = 0), implying an isotropic energy release of 
1.15$\times 10^{53}$ erg which may be relaxed by beaming by a factor 
$>$ 10$^{2}$.
Precise astrometry indicates that the GRB coincides within 
0.2$^{\prime\prime}$ with the host galaxy, thus given support to
a massive star origin. The absolute magnitude of the 
galaxy is M$_{B}$ = $-$18.2, well below the knee of the galaxy luminosity 
function and we derive a star-forming rate of (11.5 $\pm$ 7.1) 
M$_{\odot}$ yr$^{-1}$, which is much larger than the present-day rate in 
our Galaxy. The quasi-simultaneous broad-band 
photometric spectral energy distribution of the afterglow is determined 
$\sim$ 3.5 day after the burst (Dec 12.0) implying a cooling frequency
$\nu_c$ below the optical band, i.e. supporting a jet model with 
$p$ = $-$2.30 as the index of the power-law electron distribution.  

\keywords{Gamma rays: bursts - Galaxies: general - Cosmo\-logy: observations}

\end{abstract}

%

\section{Introduction}

Gamma--ray bursts (GRBs) are flashes of cosmic high energy 
($\sim 1$~keV--10~GeV) photons (Fishman and Meegan 1995). For many years 
they remained without any satisfactory explanation since their discovery 
in 1967, but with the advent of the Italian--Dutch X--ray 
satellite BeppoSAX, it became possible to conduct deep counterpart 
searches only a few hours after a burst was detected. This led to the 
first detection of X-ray and optical afterglow for GRB 970228 
(Costa et al. 1997, van Paradijs et al. 1997) and the determination 
of the cosmological distance scale for the bursts on the basis of the 
first spectrosco\-pic measurements taken for GRB 970508, implying $z$ $\geq$ 
0.835 (Metzger et al. 1997). 

Subsequent observations in 1997-2000 have 
shown that a\-bout a third of the well localized GRBs can be associated with 
optical emission that gradually fades away over weeks to months.  
Now it is widely accepted that long duration GRBs originate at cosmological
distances with energy releases of 10$^{51}$--10$^{53}$ ergs. 
The observed afterglow satisfies the predictions of the 
"stan\-dard" relativistic fireball model, and the central engines that power 
these extraordinary events are thought to be the collap\-se of massive stars
(see Piran (1999) and van Paradijs et al. (2000) for a review).

The detection of GRB host galaxies is most essential in order
to understand the nature of hosts (morphology, star forming rates) and 
to determine the energetics of the bursts (redshifts) and offsets with
respect to the galaxy centres. About 25 hosts galaxies have been detected
so far, with redshifts $z$ in the range 0.43-4.50 and star-forming rates
in the range 0.5-60 M$_{\odot}$ year$^{-1}$. See Klose (2000),
Castro-Tirado (2001) and references therein.

Here we report on the detection of the optical afterglow from GRB 991208
as well as its host galaxy. This GRB was detected at 04:36 
universal time (UT) on 8~Dec~1999, with 
the Ulysses GRB detector, the Russian GRB Experiment 
(KONUS) on the Wind spacecraft and the Near Earth Asteroid Rendezvous (NEAR) 
detectors (Hurley et al. 2000) as an extremely intense, 60 s long GRB with a 
fluence $>$ 25 keV of $10^{-4}$ erg cm$^{-2}$ and considerable flux 
above 100 keV. 
Radio observations taken on 1999 December 10.92 UT with the Very Large Array 
(VLA) at 4.86 GHz and 8.46 GHz indicated the presence of a compact source  
which became a strong candidate for the radio afterglow from GRB 991208 
(Frail et al. 1999).

\section{Observations and data reduction}

We have obtained optical images centered on the GRB 
location starting 2.1~days after the burst (Table~1).
Photometric observations were conducted with the 1.04-m Sampurnanand 
telescope at Uttar 
Pradesh State Observatory, Nainital, India (1.0 UPSO); the 1.2-m Schmidt
telescope at Tautenburg, Germany (1.2 TBG);
the 1.5-m telescope at Observatorio de Sierra Nevada (1.5 OSN), Granada, 
Spain;
the 2.5-m Isaac Newton Telescope (2.5 INT), the 2.56-m Nordic Optical 
Telescope (NOT), the 3.5-m Telescopio Nazionale Galileo (3.5 TNG) and the 
4.2-m William Herschel Telescope (4.2 WHT) at Observatorio del Roque de los 
Muchachos, La Palma, Spain;
the 1.23-m, 2.2-m and 3.5-m telescopes at the German-Spanish Calar Alto 
Observatory (1.2, 2.2 and 3.5 CAHA respectively), Spain;  
the 3.5-m telescope operated by the Universities of Wisconsin, Indiana, 
Yale and the National Optical Astronomical Observatories (3.5-m WIYN) at 
Kitt Peak, USA; and the 6.0-m telescope at the Special Astrophisical 
Observatory of the Russian Academy of Sciences in Nizhnij Arhyz, Russia.

For the optical images, photometry was performed by means of 
SExtractor (Bertin and Arnouts 1996), making use of the corrected 
isophotal magnitude, which is appropriate for star-like objects. The DAOPHOT
(Stetson 1987) profile-fitting technique was used for the magnitude 
determination on the later epoch images, when the source is much fainter. 
Zeropoints, atmospheric extinction and color terms were computed using
observations of standard fields taken throughout the run. Magnitudes of the 
secondary standards in the GRB fields agree, within the 
uncertainties, with those given in Henden (2000). 
Zeropoint uncertainties are also included in the given errors. 

Prompt follow up spectroscopy of the OA was attempted at several telescopes 
(Table 2), but we only got a reasonable good signal-to-noise ratio (S/N) at 
the 6-m telescope SAO RAS using an integral field spectrograph MPFS 
(Dodonov et al. 1999a). One 2700-sec spectrum and one 4500-sec
spectrum were obtained on 13 and 14 Dec 1999 UT. On the latter, the 
observing conditions were good: the seeing was $\sim$ 1.5$^{''}$ (at a 
zenithal distance of 60$^{\circ}$), and there was good transparency. 
We used 300 lines/mm grating blazed at 6000 \AA \rm \ giving a spectral 
resolution of about 5 \AA/pixel and effective wave\-length coverage of 
4100 - 9200 \AA. The spectrophotometric standards HZ44 and 
BD+75$^{\circ}$325 (Oke et al. 1995) were used for the flux calibration.

\begin{table}[t]
\begin{center}
\caption{Journal of the  GRB~991208 optical/NIR observations}
\begin{scriptsize}
\begin{tabular}{ccccc}
\hline
Date of     & Telescope & Filter & Integration & Magnitude \\
1999 (UT)   &           &        & time (s)    &           \\
\hline
\noalign{\smallskip}
{\small 10.2708 Dec}  & {\small 2.5 NOT}    & {\small R} &   {\small 300}     &  {\small 18.7 $\pm$ 0.1}  \\
{\small 10.2917 Dec}  & {\small 2.5 INT}    & {\small I} &   {\small 240}     &       {\small $>$ 15.5}   \\
{\small 11.2111 Dec}  & {\small 1.3 TBG}    & {\small I} &   {\small 900}     & {\small 18.75 $\pm$ 0.11} \\
{\small 11.2111 Dec}  & {\small 2.2 CAHA}   & {\small R} &   {\small 600}     & {\small 19.60 $\pm$ 0.03} \\
{\small 11.2507 Dec}  & {\small 2.2 CAHA}   & {\small R} &   {\small 600}     & {\small 19.61 $\pm$ 0.04} \\
{\small 11.2792 Dec}  & {\small 2.5 INT}    & {\small R} &   {\small 300}     & {\small 19.70 $\pm$ 0.08} \\
{\small 11.2833 Dec}  & {\small 2.5 INT}    & {\small I} &   {\small 300}     &  {\small 19.2 $\pm$ 0.1}  \\
{\small 12.0208 Dec}  & {\small 1.0 UPSO}   & {\small I} &  {\small 2 x 200}  &  {\small 19.9 $\pm$ 0.3}  \\
{\small 12.2000 Dec}  & {\small 1.2 CAHA}   & {\small B} &   {\small 300}     &        {\small $>$  20.3} \\
{\small 12.2056 Dec}  & {\small 1.2 CAHA}   & {\small V} &   {\small 300}     &        {\small $>$  20.5} \\
{\small 12.2181 Dec}  & {\small 1.2 CAHA}   & {\small R} &   {\small 300}     &  {\small 20.0 $\pm$ 0.3}  \\
{\small 12.2229 Dec}  & {\small 1.2 CAHA}   & {\small V} &   {\small 500}     &  {\small 20.7 $\pm$ 0.4}  \\
{\small 12.2299 Dec}  & {\small 1.2 CAHA}   & {\small B} &   {\small 500}     &  {\small 21.3 $\pm$ 0.2}  \\
{\small 12.2500 Dec}  & {\small 1.5 OSN}    & {\small R} &  {\small 2 x 600}  & {\small 19.9 $\pm$ 0.5}  \\
{\small 12.2535 Dec}  & {\small 1.2 CAHA}   & {\small U} &  {\small 6 x 500}  &  {\small $>$ 19.8}  \\
{\small 12.2576 Dec}  & {\small 1.5 OSN}    & {\small I} &   {\small 300}     &  {\small 19.8 $\pm$ 0.5}  \\
{\small 12.2604 Dec}  & {\small 2.5 NOT}    & {\small R} & {\small 3 x 300}   & {\small 20.37 $\pm$ 0.05} \\
{\small 12.2694 Dec}  & {\small 2.5 NOT}    & {\small I} &  {\small 3 x 300}  & {\small 19.95 $\pm$ 0.05} \\
{\small 12.2757 Dec}  & {\small 2.5 INT}    & {\small B} &   {\small 500}     & {\small 21.40 $\pm$ 0.05} \\
{\small 12.2792 Dec}  & {\small 2.5 NOT}    & {\small V} &   {\small 300}     & {\small 20.85 $\pm$ 0.05} \\
{\small 12.2806 Dec}  & {\small 2.5 INT}    & {\small V} &   {\small 300}     & {\small 20.78 $\pm$ 0.06} \\
{\small 12.2840 Dec}  & {\small 2.5 INT}    & {\small I} &   {\small 180}     & {\small 20.00 $\pm$ 0.17} \\
{\small 12.2882 Dec}  & {\small 3.5 TNG}    & {\small R} &   {\small 500}     &  {\small 20.0 $\pm$ 0.3}  \\
{\small 13.0000 Dec}  & {\small 1.0 UPSO}   & {\small I} & {\small 3 x 600}   &  {\small 20.3 $\pm$ 0.2}  \\
{\small 13.2604 Dec}  & {\small 2.5 NOT}    & {\small R} &  {\small 3 x 300}  & {\small 20.89 $\pm$ 0.04} \\
{\small 13.2715 Dec}  & {\small 2.5 INT}    & {\small B} &   {\small 500}     & {\small 22.03 $\pm$ 0.06} \\
{\small 13.2729 Dec}  & {\small 2.5 NOT}    & {\small I} & {\small 3 x 300}   & {\small 20.34 $\pm$ 0.06} \\
{\small 13.2764 Dec}  & {\small 2.5 INT}    & {\small V} &   {\small 180}     & {\small 21.36 $\pm$ 0.07} \\
{\small 13.2799 Dec}  & {\small 2.5 INT}    & {\small I} &   {\small 300}     & {\small 20.26 $\pm$ 0.11} \\
{\small 13.2833 Dec}  & {\small 2.5 NOT}    & {\small V} &   {\small 300}     & {\small 21.38 $\pm$ 0.07} \\
{\small 13.2910 Dec}  & {\small 3.5 TNG}    & {\small R} &   {\small 360}     &  {\small 20.8 $\pm$ 0.3}  \\
{\small 14.2708 Dec}  & {\small 2.5 NOT}    & {\small R} &  {\small 3 x 300}  & {\small 21.43 $\pm$ 0.04} \\
{\small 14.2743 Dec}  & {\small 2.5 INT}    & {\small B} &  {\small 1,000}    & {\small 22.31 $\pm$ 0.08} \\
{\small 14.2778 Dec}  & {\small 2.5 INT}    & {\small U} &   {\small 535}     &        {\small $>$ 23.0}  \\
{\small 14.2792 Dec}  & {\small 2.5 NOT}    & {\small I} &  {\small 3 x 300}  & {\small 20.91 $\pm$ 0.07} \\
{\small 14.2847 Dec}  & {\small 3.5 TNG}    & {\small R} &   {\small 600}     & {\small 21.40 $\pm$ 0.10} \\
{\small 14.2875 Dec}  & {\small 2.5 NOT}    & {\small V} &   {\small 300}     & {\small 21.68 $\pm$ 0.10} \\
{\small 15.2708 Dec}  & {\small 2.5 NOT}    & {\small R} &  {\small 3 x 300}  & {\small 21.97 $\pm$ 0.08} \\
{\small 15.2833 Dec}  & {\small 2.5 NOT}    & {\small I} &  {\small 3 x 300}  & {\small 21.46 $\pm$ 0.16} \\
{\small 15.2938 Dec}  & {\small 2.5 NOT}    & {\small V} &  {\small 2 x 300}  &       {\small $>$21.8}    \\
{\small 03.5319 Jan}  & {\small 3.5 WIYN}   & {\small R} &   {\small 600}     &        {\small $>$ 23.0}  \\
{\small 03.5507 Jan}  & {\small 3.5 WIYN}   & {\small I} &   {\small 600}     &        {\small $>$ 22.0}  \\
{\small 04.2292 Jan}  & {\small 2.2 CAHA}   & {\small R} &  {\small 2 x 900}  &        {\small $>$ 23.5}  \\
{\small 05.2292 Jan}  & {\small 3.5 CAHA}   & {\small R} & {\small 2 x 1,200} & {\small 23.23 $\pm$ 0.13} \\
{\small 06.2083 Jan}  & {\small 2.2 CAHA}   & {\small V} & {\small 9 x 1,200} & {\small 23.83 $\pm$ 0.10} \\
{\small 13.2097 Jan}  & {\small 3.5 CAHA}   & {\small R} & {\small 3 x 1,200} &        {\small $>$ 23.1}  \\
{\small 13.2528 Jan}  & {\small 3.5 CAHA}   & {\small V} & {\small 2 x 1,200} &        {\small $>$ 22.5}    \\
{\small 19.2604 Jan}  & {\small 2.5 NOT}    & {\small R} & {\small 7 x 600}   & {\small 23.65 $\pm$ 0.13} \\
{\small 29.2431 Jan}  & {\small 4.2 WHT}    & {\small I} &  {\small 2 x 900}  &        {\small $>$ 22.5}  \\
{\small 29.2604 Jan}  & {\small 4.2 WHT}    & {\small B} &   {\small 986}     &        {\small $>$ 23.6}  \\    
{\small 13.2556 Feb}  & {\small 3.5 TNG}    & {\small B} & {\small 3 x 1,200} & {\small 24.65 $\pm$ 0.04} \\
{\small 17.2882 Feb}  & {\small 3.5 TNG}    & {\small V} & {\small 2 x 1,200} & {\small 24.22 $\pm$ 0.09} \\
{\small 31.8403 Mar}  & {\small 6.0 SAO}    & {\small V} &   {\small 1,490}   & {\small 24.55 $\pm$ 0.16} \\
{\small 31.8715 Mar}  & {\small 6.0 SAO}    & {\small I} &    {\small 360}    & {\small 23.46 $\pm$ 0.49} \\
{\small 31.9028 Mar}  & {\small 6.0 SAO}    & {\small B} &   {\small 1,795}   & {\small 25.19 $\pm$ 0.17} \\
{\small 31.9583 Mar}  & {\small 6.0 SAO}    & {\small R} &   {\small 1,260}   & {\small 24.27 $\pm$ 0.15} \\
{\small 04.2083 Apr}  & {\small 2.5 NOT}    & {\small I} &   {\small 3,800}   & {\small 23.3 $\pm$ 0.2} \\
\hline
{\small 11.2708 Feb}  & {\small 3.5 TNG}    & {\small J} &  {\small 42 x 60}  &       {\small $>$ 22.0}   \\
\noalign{\smallskip}
\hline
\end{tabular} 
\end{scriptsize}
\end{center}
\end{table}

\noindent
\begin{table}[t]
\begin{center}
\caption{Journal of the GRB~991208 spectroscopic observations}
\begin{tabular}{cccc}
\noalign{\smallskip}
\hline
Date of     & Telescope & Wavelength range & Exposure   \\
1999 (UT)   &           & (A)              & time (s)   \\
\noalign{\smallskip}
\hline
\noalign{\smallskip}
12.2306 Dec  & 2.2 CAHA    & 3,550--4,510 &   1,800     \\
12.2347 Dec  & 3.5 CAHA    & 6,000--10,000&   1,800     \\
13.2083 Dec  & 6.0 SAO     & 4,100--9,200 &   2,700     \\
14.2083 Dec  & 6.0 SAO     & 4,100--9,200 &   4,500     \\
18.2431 Dec  & 4.2 WHT     & 4,000--9,000 &   3,600     \\
\noalign{\smallskip}
\end{tabular}
\end{center}
\end{table}

\begin{figure}[H]
\begin{center}
  \resizebox{7cm}{7cm}{\includegraphics{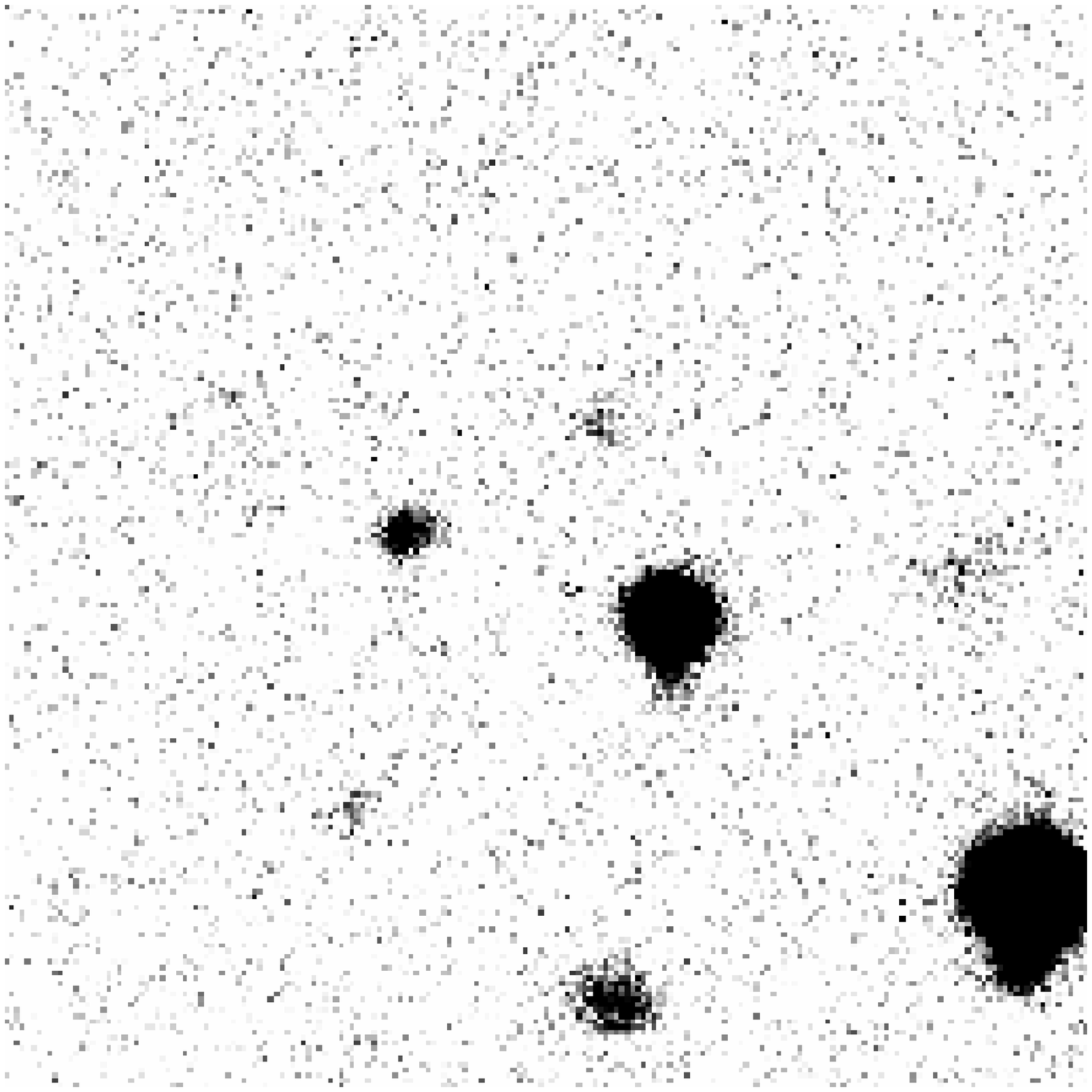}}
  \resizebox{7cm}{7cm}{\includegraphics{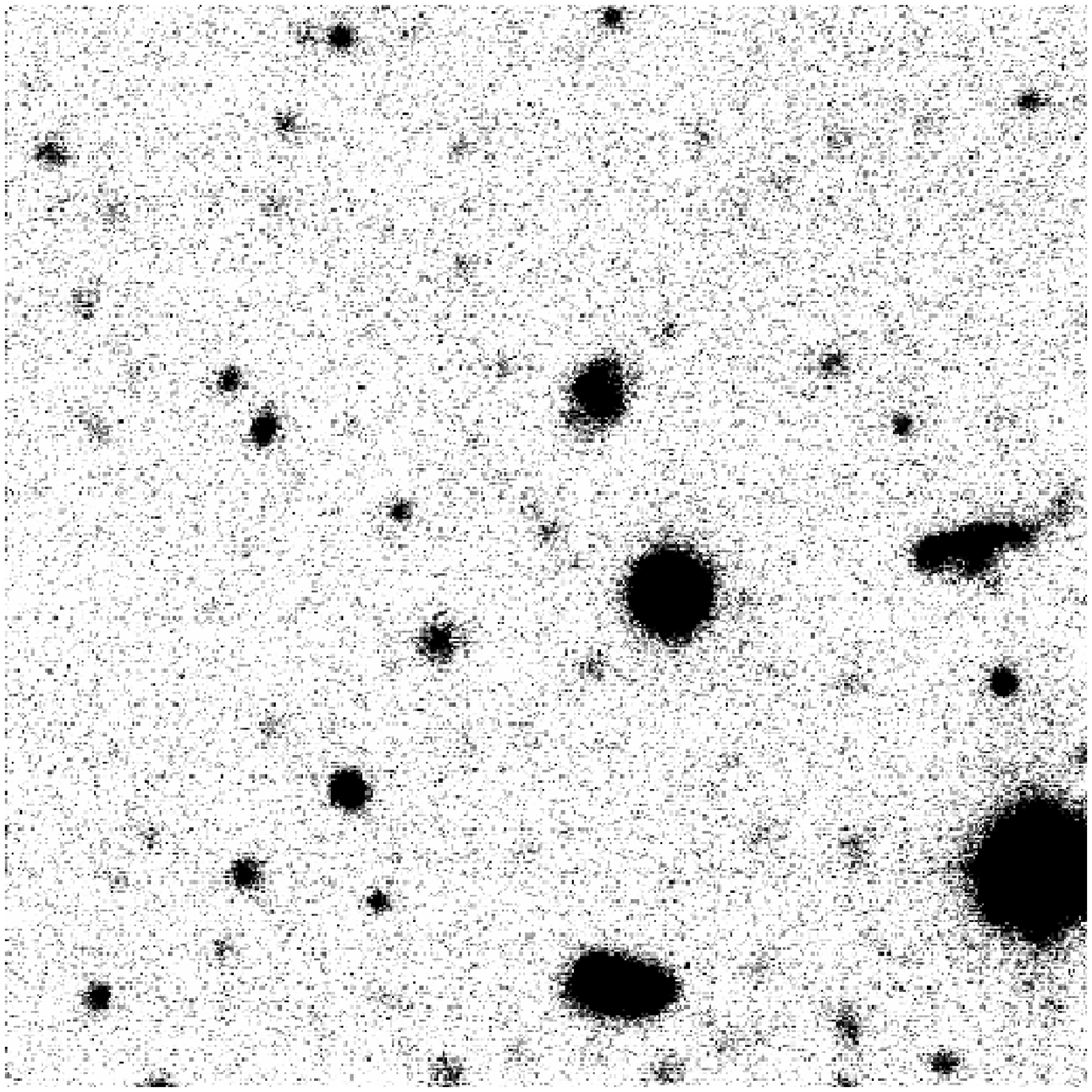}}
    \caption{Blue (B band) images of the GRB 991208 location. The frames were 
             taken at the 2.5-m INT on 12 Dec 1999 (a: {\it upper 
             panel}, 3.9 d after the GRB), and at the 3.5-m TNG on 13 
             Feb 2000 (b: {\it left panel}, 36 days after the GRB ). 
             It shows the optical afterglow and the underlying galaxy close to 
             the center of the image. Here only a 1.$^{'}1 \times 1.^{'}$1 
             field of view is presented. The positions of both objects are 
             consistent within the astrometric uncertainty (0.2$^{''}$). 
             North is at the top and east to the left. Limi\-ting magnitudes 
             were B $\sim$ 22.5 and B $\sim$ 25.5, respectively.}
\label{figure1}
\end{center}
\end{figure}

\section{Results and discussion}

\subsection{The optical afterglow}

At the same location of the variable radiosource, a bright optical afterglow 
(OA) was identified on the images taken at Calar Alto, La Palma and
Tautenburg (Castro-Tirado et al. 1999a,b). The astrometric solution was 
obtained using 16 USNO-A stars, and coordinates were
$\alpha_{2000}=16^h 33^m 53.^s50; \delta_{2000}=+46^{\circ} 27^{'} 21.0^{''}$
($\pm 0.2^{''}$).  A comparison among optical images acquired on 
10 and 11~Dec allowed us to confirm the variability in intensity
of the proposed OA. 
About 2.1 d after the burst, we measured R = 18.7 $\pm$ 0.1 for the OA, 
and 19 h later we found R = 19.60 $\pm$ 0.03. 
In these images the object is point-like (resolution $\sim$ 
1$^{\prime\prime}$) and there is no evidence of any underlying extended 
object as seen at later epochs (Fig.~1).
Coincident (within errors) with the location of 
optical and radio afterglows, Shepherd et al. (1999) detected at millimeter 
wavelengths the brightest afterglow of a GRB reported so far. 
At 15 GHz and 240 GHz, the GRB 991208 afterglow was observed at Ryle
(Pooley et al. 1999) and Pico Veleta (Bremer et al. 1999a,b), respectively. 

Our $B,V,R,I$ light curve (Fig. 2) shows that the source was declining
in brightness.
The optical decay slowed down in early 2000, indicating the presence
of an underlying source of constant brightness: the host galaxy.
The decay of previous GRB afterglows appears 
to be well characterized by a power law (PL) decay 
$F(t) \propto (t-t_0)^{\alpha}$, 
where $F(t)$ is the flux of the afterglow at time $t$ since the onset of
the event at $t_{0}$ and $\alpha$ is the decay
constant.  Assuming this parametric form and by fitting least square linear 
regressions to the observed magnitudes as function of time, we derive below
the value of flux decay constant for GRB 991208. 
The fits to the $B$, $V$, $R$ and $I$ light curves are given in Table 3, 
but the poor quality of the PL fit is reflected in the relatively
large reduced chi-squared values. This is specially noticeable in
the R-band light curve, due to the data obtained after one month,
that will be discussed in Section 3.1.2.

\subsubsection{The existence of two breaks}

The $R$ and $I$-band data up to 
$t_{0}$ + 10 days are better fit by a broken PL with a 
break time $t_{break}$ $\sim$ 5 days. For the B and V-band such a
fit is not possible due to the scarcely of the data in these bands.
See Table 4. Hence we adopt a value of $\alpha_1$ = $-$2.30$\pm$0.07 for 
2 days $<$ $(t-t_0)$ $<$ 5 days and $\alpha_2$ = $-$3.2$\pm$0.2 for 
5 days $<$ $(t-t_0)$ $<$ 10 days
as flux decay constants in further discussions.

Further support for the existence of an additional break at  $(t-t_0)$ $<$ 2 
days in GRB 991208 comes from
the extrapolation of the R-band data towards earlier epochs 
(Fig. 3), that pre\-dicts an optical flux that should have been seen at naked 
eye by observers in Central Europe.

\noindent
\begin{table}[H]
\begin{center}
\caption{PL fits to the BVRI observations of GRB 991208} 
\begin{tabular}{ccc}
\hline~
Filter &     $\alpha$       & $\chi^{2}/$dof \\
\noalign{\smallskip}
\hline
\noalign{\smallskip}
  B    &   $-1.37 \pm 0.04$ &  24.6/3    \\
  V    &   $-1.75 \pm 0.07$ &  16.0/6    \\
  R    &   $-2.22 \pm 0.04$ &  91.5/11   \\ 
  I    &   $-2.58 \pm 0.12$ &  19.3/8    \\
\noalign{\smallskip}
\hline
\end{tabular}
\end{center}
\end{table}

\noindent
\begin{table}[H]
\begin{center}
\caption{Broken PL fits to the RI observations of GRB 991208} 
\begin{tabular}{ccccc}
\hline~
Filter & $\alpha_1$ & $\chi^{2}/$dof &  $\alpha_2$  & $\chi^{2}/$dof\\
\noalign{\smallskip}
\hline
\noalign{\smallskip}
  R    &   $-2.30 \pm 0.07$ &   3.9/6   &   $-3.18 \pm 0.22$ &   5.7/3  \\ 
  I    &   $-2.51 \pm 0.27$ &   0.4/3   &   $-3.33 \pm 0.39$ &   1.1/3  \\
\noalign{\smallskip}
\hline
\end{tabular}
\end{center}
\end{table}

\begin{figure}
\resizebox{9.0cm}{9.0cm}{\includegraphics[angle=0]{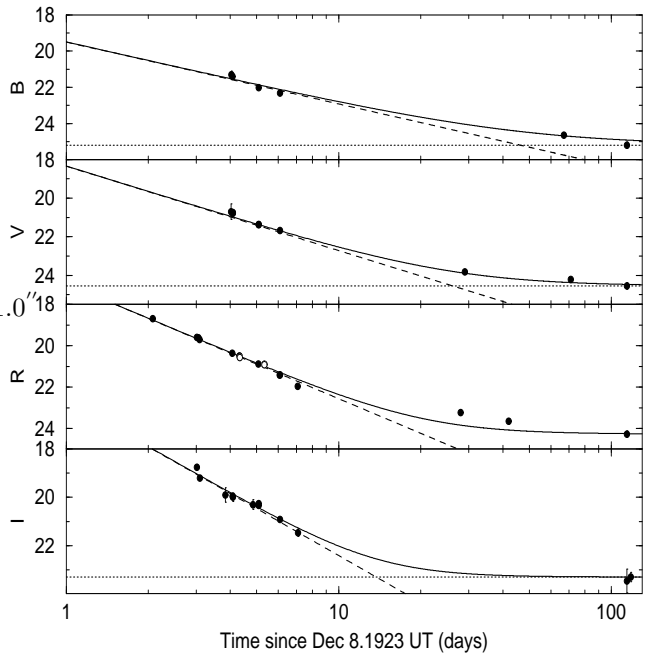}}
    \caption{The $BVRI$-band light-curves of the optical transient related 
             to GRB 991208, including the underlying galaxy. 
             Filled circles are our data are empty circles are data from
             Garnavich and Noriega-Crespo (1999) and Halpern and 
             Helfand (1999). The dashed-line is the pure OA contribution 
             to the total flux, according to the single power-law fits given 
             on Table 2. The dotted line is the contribution of the host 
             galaxy. The solid line is the combined flux (OA plus underlying 
             galaxy).}
\label{figure2}
\end{figure}

The optical event exceeding magnitude 11 could be detected by the Czech 
stations of the European Fireball Network. Unfortunately, it was completely 
cloudy during the night of Dec 8/9 in the Czech Republic, so none of the 
12 stations of the network was able to take all-sky photographs. The first 
photographs after the GRB trigger were taken on Dec 8, 16:25 UT, i.e. nearly 
12 h after the event, and shows no object at its position brighter than mag 
$V \sim$ 10. However, sky patrol films taken for meteor 
research were exposed in Germany during Dec 8/9, 1999 but no OA exceeding 
$R \sim$ 4 with a duration of 10 s or more is detected simultaneously to the 
GRB event. 
This upper limit derived from the films implies that this additional break 
in the power-law decay of GRB 991208 has to be present at 0.01 days 
$<$ $(t-t_0)$ $<$ 2 days although a maximum in the light curve similar to 
GRB 970508 (Castro-Tirado et al. 1998) cannot be excluded.

The flux decay of  GRB 991208 is one of the steepest of all GRBs observed so 
far (Sagar et al. 2000). Before deriving any conclusion from the flux decays 
of these GRBs, we compare them with other well studied GRBs. Most OAs 
exhibit a single power-law decay index, generally $\sim$ $-$ 1.2, 
a value reasonable for spherical expansion of a relativistic blast wave 
in a constant density interstellar medium (M\'esz\'aros and Rees 1997, 
Wijers et al. 1997, Waxman 1997, Reichart 1997) . 
For other bursts, like GRB 990123, the value of $\alpha$ = $-$1.13$\pm$0.02 
for the early time (3 hr to 2 day) light curve becomes $-$1.75$\pm$0.11 at 
late times (2-20 day) 
(Kulkarni et al. 1999, Castro-Tirado et al. 1999c, Fruchter et al. 1999)
while the corresponding slopes for GRB 990510 are $-$0.76$\pm$0.01 and 
$-$2.40$\pm$0.02 respectively with the $t_{break}$ $\sim$ 1.57 day 
(Stanek et al. 1999, Harrison et al. 1999). 
If the steepening observed in both cases is due to beaming, then one may 
conclude that it occurs within $<$ 2 days of the burst. 

Rapid decays in OAs have been seen in GRB 980326 with
$\alpha$ = $-$2.0$\pm$0.1 (Bloom et al. 1999a),
GRB 980519 with $\alpha$ = $-$2.05$\pm$0.04 (Halpern et al. 1999), 
GRB 990510 with $\alpha$ = $-$2.40 $\pm$ 0.02 (Stanek et al. 1999, Harrison et
al. 1999) and GRB 000301C with $\alpha$ = $-$2.2$\pm$ 0.1 (Masetti et al. 
2000, Jensen et al. 2001, Rhoads and Fruchter 2001), and have been 
interpreted by the
sideways expansion of a jet (Rhoads 1997, 1999, M\'esz\'aros and Rees 1999).
For GRB 991208, $\alpha_1$ = $-$2.30$\pm$0.07 and we therefore argue 
that the observed steep decay in the optical light curve up to $\sim$ 5 days 
may be due to a break which occurred before our first optical 
observations starting $\sim$ 2.1 day after the burst. The break is 
expected in several physical models, but beaming is the most likely 
cause in GRB 991208 taking into account that the rapid 
fading of optical afterglows is conside\-red as an evidence for beaming 
in GRBs (Huang, Dai and Lu 2000). 

\begin{figure}
\resizebox{9.5cm}{10cm}{\includegraphics[angle=0]{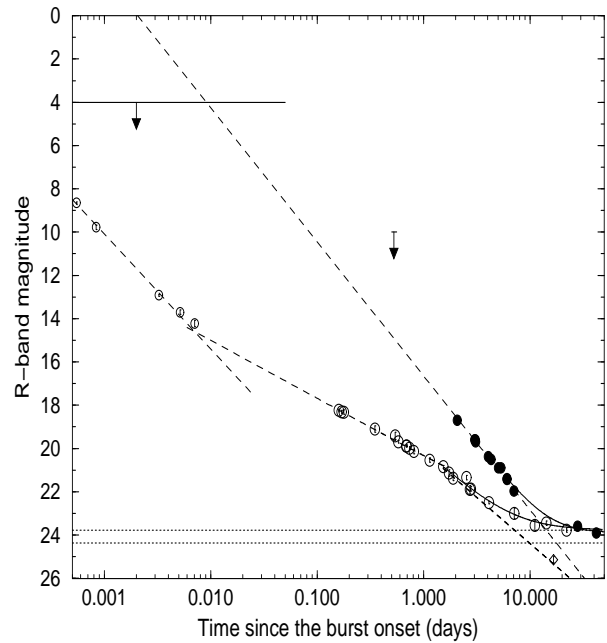}}
    \caption{Comparison of the two brightest optical GRB afterglows detected 
             so far: GRB 990123 (empty circles, from Castro-Tirado et al. 
             1999c) and GRB 991208 (filled circles, from this paper). 
             The extrapolation of the GRB 991208 R-band data towards earlier 
             epochs predicts an optical flux that should have been seen at 
             naked eye by obser\-vers in Central Europe. However, the upper 
             limit (R $\sim$ 4) derived from simultaneous sky patrol films 
             implies that either a break in the power-law decay or a maximum 
             in the light curve has to be present at 0.01 $<$ T $<$ 2 days. 
             The dotted lines are the constant contribution of the two host 
             galaxies, R $\sim$ 23.9 and 24.3 respectively. The dashed-lines 
             are the pure OAs contributions to the total fluxes. The solid 
             lines (only shown here for clarity after T $>$ 5 days) are the 
             combined fluxes (OA plus underlying galaxy on each case).}
\label{figure3}
\end{figure}

According to the current view, the forward external shock wave would have 
led to the afterglow as observed in all wavelengths.
The population of electrons is assumed to be a power-law 
distribution of Lorentz factors $\Gamma_{e}$ following d$N$/d$\Gamma_{e}$ 
$\propto$ $\Gamma_{e}^{-p}$ above a minimum Lorentz factor
$\Gamma_{e}$ $\geq$ $\Gamma_{m}$, corresponding to the 
synchrotron frequency $\nu_{m}$. 
The value of $p$ can be determined taking into account the occurrence 
of the {\it jet effect}: the break due to a lateral expansion in the 
decelerating jet occurs when the initial Lorentz factor 
$\Gamma$ drops below $\theta_{0}$$^{-1}$ (with $\theta_{0}$ the initial 
opening angle), i.e. the observer ``sees'' the edge of the jet.  
A change in the initial power-law decay exponent $\alpha_0$ (unknown to us) 
from $\alpha_0$ = $3(1-p)/4$ to $\alpha_1$ = $-p$  
(for $\nu_{m}$ $<$ $\nu$ $<$ $\nu_{c}$), 
or from $\alpha_0$ = $(2-3p)/4$ to $\alpha_1$ = $-p$ 
(for $\nu$ $\geq$ $\nu_{c}$) is expected (Rhoads 1997, 1999).  
If this is the case, then $p$ = $-\alpha_1$ = 2.30 $\pm$ 0.07, 
in the observed range for other GRBs.

\begin{figure*}
\resizebox{17cm}{13cm}{\includegraphics[angle=0]{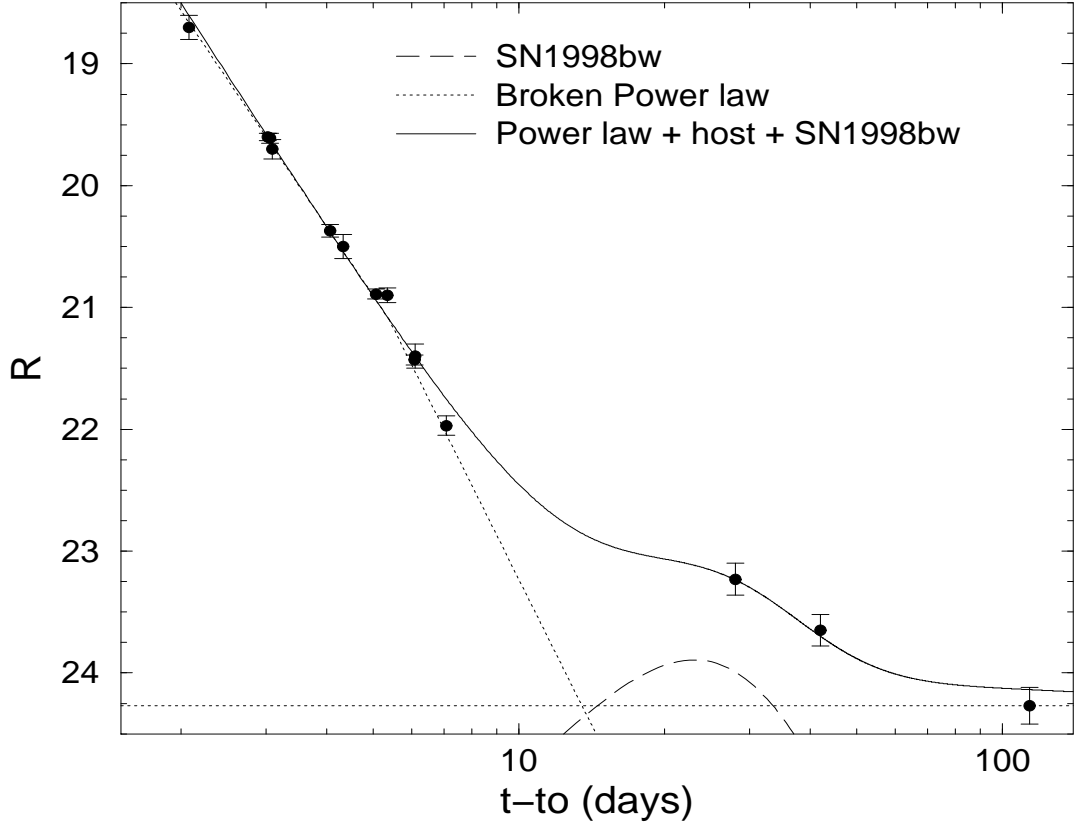}}
    \caption{The GRB 991208 R-band light curve (solid line) fitted with 
            a SN1998 bw-like component at $z$ = 0.706 (long dashed line) 
            superposed to the broken power-law OA light curve displaying
            the second break at t$_{break}$ $\sim$ 5 $d$ 
            (with $\alpha_1$= $-$2.3 and $\alpha_2$ = $-$3.2, 
            short dotted lines) 
            and the constant contribution of the host galaxy 
            (R = 24.27 $\pm$0.15, dotted line).}
\label{figure4}
\end{figure*}

Whether the jet was expanding into an constant density medium or
in an inhomogeneus medium (Chevalier and Li 1999, Wei and Lu 
2000) is not possible to know with our data alone, as we do not 
have information on $\alpha_0$. For a density gra\-dient of $s$ = 2, 
as expected from a previously ejected stellar wind ($\rho \propto r^{-s}$), 
the light curve should steepen by $\Delta\alpha$ = ($\alpha_1-\alpha_0$) 
= (3-$s$)/(4-$s$) = 0.5 whereas $\Delta\alpha$ = 0.75 for a constant 
density medium. 
 
But which is the reason of the second break observed in GRB 991208 after 
$\sim$ 5 days ? The passage of the cooling frequency $\nu_{c}$ through the 
optical band (that would steepen the light curve by $\Delta\alpha$ $\sim$ 
0.25, Sari et al. 1998) can be discarded:  
fo\-llowing Sari, Piran \& Halpern (1999), 
if $\nu < \nu_{c}$ then we should expect a spectral index $\beta$ 
($F_\nu \propto \nu^{-\beta}$) such as $\beta$ = (p$-$1)/2 = 0.65 $\pm$ 0.04
and if $\nu > \nu_{c}$ then $\beta$ = p/2 = 1.15 $\pm$ 0.04 which is 
compatible with $F_\nu \propto \nu^{-1.05 \pm 0.08}$ on 12 Dec 
(see Section 3.3). 
Hence we conclude that $\nu_{c}$ has already passed the optical band 4 days 
after the burst onset.

The difference between the mid and late time decay slopes is
 $\Delta\alpha$ = ($\alpha_2-\alpha_1$) = 0.9 $\pm$ 0.3. 
A possible explanation could be two superposed events: a major burst
followed by a minor burst, expected from some SN-shock models 
(M\'eszaros, Rees and Wijers 1998) similarly to what has been proposed 
for GRB 000301C (Bhargavi and Cowsik 2000). 
Li and Chevalier (2001) find that a 
spherical wind model (with $\rho \propto r^{-2}$) and a jet model fit
the radio data  when using a steepening of the electron spectrum, invoking 
a non-standard, broken PL around a break Lorentz factor $\Gamma_{break}$: 
$dN_e/d\Gamma$ = $C_1 \Gamma^{-p_1}$ if $\Gamma_{min} < \Gamma 
< \Gamma_{break}$ and $dN_e/d\Gamma$ = 
$C_2 \Gamma^{-p_2}$ if $\Gamma > \Gamma_{break}$. 
They derive $p_1$ = 2.0 and $p_2$ = 3.3, being the later value
consistent with $-\alpha_2$.

\subsubsection{The late--time light curve: another underlying SN?}

If an underlying supernova (SN) would be present in the GRB 991208
light curve, this is expected to peak at $\sim$ 15(1+$z$) days $\sim$ 25 days.
GRB 990128 is a good candidate for such a search thanks to the rapid
decay.
Indeed, the late--time light curve in the optical band (specially in the
R-band) cannot be acceptably fitted just with the power-law decline 
expected for the OA plus the constant contribution of the host galaxy
($\chi^2$/dof = 8.32). The data is much better fitted when considering
a third component, a type Ic SN1998bw-like component (Galama et al. 1998) 
at $z$ = 0.706 ($\chi^2$/dof = 1.88). See Fig. 4. We have used SN1998bw
because its likely relationship to GRB 980425.

Thus, GRB 991208 would be the sixth event
for which contribution from a SN is proposed, after GRB 970228 (Reichart 
1999, Galama et al. 2000b), GRB 970508 (Sokolov et al. 2001a),
GRB 980326 (Castro-Tirado and Gorosabel 1999, Bloom et al. 1999a), 
GRB 990712 (Hjorth et al. 1999, Sahu et al. 2000) 
and GRB 000418 (Klose et al. 2000, Dar and De R\'ujula 2001). 
This reinforces the GRB-SN relationship for some long duration
bursts and support the scenario in which the death of a massive star
produces the GRB  in the ``collapsar'' model (MacFadyen \& 
Woosley 1999). Our results do not support the ``supranova'' model 
(Vietri \& Stella 1998) for this event as the SN should have preceeded 
the GRB by few months.

But could the observations be explained by a dust echo instead ? 
Esin and Blandford (2000) presented an
alternative explanation for the excess of red flux observed 20-30
days after GRB 970228 and GRB 980326, being scattering off dust grains, 
peaking around $\sim$ 4000 \AA \rm ~in the rest frame (i.e. in the R-band
at $z$ = 0.706, as observed in GRB 991208). On the basis of VRIJK observations 
for GRB 970228, Reichart (2001) has concluded that the late-time afterglow
of that event cannot be explained by a dust echo. For GRB 991208, only
V- and R-band data (plus an upper limit in the I-band) are available
at the time of the maximum, i.e. the light curve is not sufficiently
well sampled to distinguish between a SN and a dust echo.

\subsection{The host galaxy}

Evidence for a brigh host galaxy came from 
the BTA/MPSF 4500-s spectrum of the GRB 991208 optical transient
taken on 14 Dec (see Fig. 5). We found four emission lines at $\lambda$ =
6350 \AA, 8300 \AA, 8550 \AA, and 8470 \AA, with the most likely 
identifications of these emission lines being: [OII] 3727 \AA,  
H$_{\beta}$ 4861 \AA, [OIII] 4959 \AA, 5007 \AA \rm \ at a redshift of 
z = 0.7063 $\pm$ 0.0017 (Dodonov et al. 1999b), 
a value confirmed by other measurements later on (Djorgovski et al. 1999). 
Line parameters are measured with Gaussian fit to the
emission line and a flat fit to the continuum.    

Considering the redshift of z = 0.7063 $\pm$ 0.0017, 
H$_{0}$ = 60 km s$^{-1}$ Mpc$^{-1}$, $\Omega_{0}$ = 1 and  $\Lambda_{0}$ = 0, 
the luminosity distance to the host is d$_{L}$ = 1.15 $\times$ 10$^{28}$ cm,
implying an isotropic energy release of 1.15 $\times 10^{53}$ erg. 
Taking into accout the time of the break, t$_{break}$ $<$ 2 d, this 
implies an upper limit on the jet half-opening angle 
$\theta_{0}$ $<$ 8$^{\circ}$ n$^{1/8}$ with $n$ the density of the ambient
medium (in cm$^{-3}$) (see Wijers and Galama 1999), and thus the energy 
release should be 
lowered by $>$ 100, i.e. the energy released is 
$<$ 1.15 $\times 10^{51}$ erg. 

For the galaxy, which is present in the late images (March-April 2000), 
the astrometric solution also obtained using the same 16 USNO-A stars was
$\alpha_{2000}=16^h 33^m 53.^s53; \delta_{2000}=+46^{\circ} 27^{'} 21.0^{''}$
($\pm 0.2^{''}$), which is consistent with the OA position. See also
Fruchter et al. (2000). 

Our broad-band measurements of B = 25.19 $\pm$ 0.17, 
V = 24.55 $\pm$ 0.16, R = 24.27 $\pm$ 0.15 on 
31.9~Mar and I = 23.3 $\pm$ 0.2  on 4.2~Apr, once 
derredened by the Galactic extinction imply a spectral 
distribution  $F_\nu$ $\propto$ $\nu^{\beta}$ with $\beta$ = 
$-$1.45 $\pm$ 0.33 ($\chi^{2}$ per degree of freedom, $\chi^{2}/$dof 
= 1.20). See Sokolov et al. (2000) for further details.
The unobscured flux density at 7510 \AA, the redshifted effective
wavelength of the B-band, is $\sim$ 0.65 $\mu$Jy, corresponding to an
absolute magnitude of M$_{B}$ = $-$18.2,  well below the knee of the
galaxy luminosity function,  M$_{B}$$^{*}$ $\sim$ $-$20.6 (Schechter 1976). 

The star-forming rate (SFR) can be estimated in different ways. Again, 
we have assumed H$_{0}$ = 60 km s$^{-1}$ Mpc$^{-1}$ and $\Omega_{0}$ = 1, 
$\Lambda_{0}$ = 0. 
From the H$_{\beta}$ flux which is (3.84 $\pm$ 0.33) x 10$^{-16}$ erg 
cm$^{-2}$ s$^{-1}$, this corresponds to (18.2 $\pm$ 0.6) M$_{\odot}$ 
yr$^{-1}$ (Pettini et al. 1998). 
From the [O II]  3727 \AA \rm \ flux (Kennicutt 1998), which is 
(1.79 $\pm$ 0.22) 
x 10$^{-16}$ erg cm$^{-2}$ s$^{-1}$ we get (4.8 $\pm$ 0.2) M$_{\odot}$ 
yr$^{-1}$. The mean value, (11.5 $\pm$ 7.1) M$_{\odot}$ yr$^{-1}$, is 
much larger than the present-day rate in our Galaxy. In any case, 
this estimate is only a lower limit on the SFR due to the unknown rest frame 
host galaxy extinction. See also Sokolov et al. (2001b).

\begin{figure}
\resizebox{\hsize}{!}{\includegraphics[angle=0]{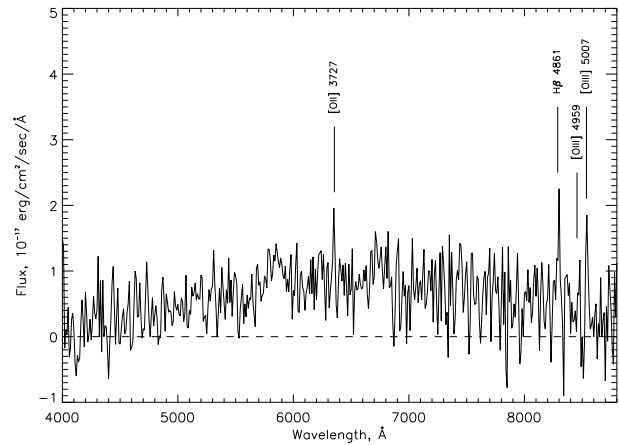}}
    \caption{The BTA/MPSF optical 4500-s spectrum of the GRB 991208 afterglow
             (on Dec 14.14, 1999 UT) in the 4000 - 8800 \AA \rm \ spectral 
             range. The spectrum has been boxcar smoothed with a 10 \AA \rm 
             ~window. The detection of the four emission lines led to a 
             redshift of z = 0.7063 $\pm$ 0.0017, that of the host galaxy.}
\label{figure5}
\end{figure}

\noindent
\begin{table}[]
\begin{center}
\caption{Journal of the GRB~991208 line identifications}
\begin{tabular}{lccc}
\noalign{\smallskip}
\hline
Line    &              Fluxes                  &  - EW         & FWHM      \\
 ID     & (10$^{-16}$ erg cm$^{-2}$ s$^{-1}$)  &   ($\rm \AA$) & ($\rm \AA$)\\
\noalign{\smallskip}
\hline
\noalign{\smallskip}
\rm [OII]  3727   & (1.79 $\pm$ 0.22) &   20  &  15.4 $\pm$ 2.0 \\
H$_{\beta}$       & (3.84 $\pm$ 0.33) &   93  &  22.3 $\pm$ 2.2 \\
\rm [OIII] 4958.9 & (1.61 $\pm$ 0.32) &   80  &  18.7 $\pm$ 4.5 \\
\rm [OIII] 5006.9 & (4.90 $\pm$ 0.33) &  244  &  20.8 $\pm$ 1.9 \\
\noalign{\smallskip}
\end{tabular}
\end{center}
\end{table}

\subsection{The multiwavelength spectrum on Dec 12.0}

We have determined the flux distribution of the GRB 991208 OA on 
Dec 12.0, 1999 UT by means of our broad-band photometric measurements 
(Dec 12.2) and other data points at mm and cm wavelengths (Dec 11.6-11.8) 
(Fig. 6). We fitted the observed flux 
distribution with a power law $F_{\nu} \propto  \nu^{\beta}$, where $F_{\nu}$
is the flux at frequency $\nu$, and $\beta$ is the spectral index. 
Optical flux at the wavelengths of $B,V,R$ and $I$ passbands has been 
derived substracting the contribution of the host galaxy and assu\-ming a
reddening E(B-V) = 0.016 (Schlegel et al. 1998).
In converting the magnitude into flux, the effective wavelengths and 
normalizations given in Bessell (1979) and Bessell and Brett (1988) were used. 
The flux densities, are 11.5, 16.7, 22.0 
and 24.2 $\mu$Jy at the effective wavelengths of $B, V, R$ and $I$ passbands,
not corrected for possible intrinsic absorption in the host galaxy.
The fit to the optical data F$_{\nu}$ $\propto$ $\nu^{\beta}$ gives 
$\beta = -1.05\pm 0.09$ ($\chi^{2}/$dof = 5.7). 
This about 2$\sigma$ above the value of $\beta = -0.77\pm 0.14$ given 
on Dec 16.6 with the Keck telescope (Bloom et al. 1999b) and 
$\beta = -1.4\pm 0.4$ for the spectral index between optical to IR 
wavelengths (that differs from the one given by Sagar et al. 2000) 
when considering $\alpha_2$. 

\begin{figure}
\resizebox{9cm}{9cm}{\includegraphics[angle=0]{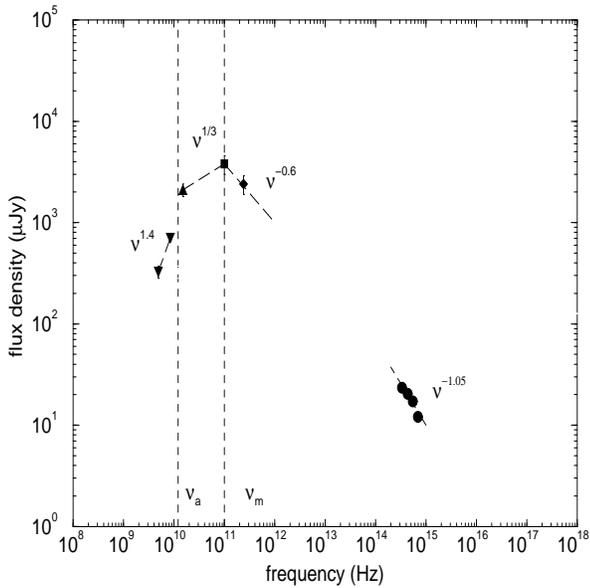}}
    \caption{The multiwavelength spectrum of GRB 991208 afterglow on 
             Dec 12.0 UT, 1999. Circles are the extrapolation of the 
             $BVRI$ measurements following the power-law derived in 
             this paper. The diamond is the Pico Veleta measurement 
             (Bremer et al. 1999a,b), 
             the square is the OVRO data point (Shepherd 1999), 
             the triangle-up is the flux density obtained at the Ryle 
             telescope (Pooley et al. 1999) and the triangles-down 
             are the VLA data points (Frail et al. 1999, Hurley et al. 
             2000). The correction for galactic extinction 
             has been considered taking into account E(B-V) = 0.016 from 
             (Schlegel et al. 1998). The long dashed lines are the fits to the 
             multiwavelength spectrum. Rough estimates of the self-absorption 
             (${\nu}_{a}$) and synchrotron frequencies (${\nu}_{m}$) are 
             also indicated.}
\label{figure6}
\end{figure}

From the maximum observed flux (Shepherd et al. 1999), we derive 
a rough value of $\nu_{m}$ $\sim$ 100 GHz. The low-frequency spectrum 
below $\nu_{m}$ ($F_{\nu} \propto  {\nu}^{1/3}$) is in agreement with the 
expected tail of the synchrotron radiation plus a self absorption 
that becomes important below a critical frequency 
$\nu_{a}$ $\sim$  13 GHz,
taking into account that $F_{\nu}$ $\propto$ $\nu^{1.4}$ in the range 
4.86-8.46 GHz (Frail et al. 1999), deviating from 
$F_{\nu}  \propto \nu^{1/3}$ as seen for 15 GHz $<$ $\nu$ $<$ 100 GHz 
(Pooley et al. 1999). 
Much more accurate estimates for $\nu_{a}$ and $\nu_{m}$ are given by
Galama et al. (2000a). 
Above $\nu_{m}$, the IRAM observations 
(Bremer et al. 1999a,b)  indicate a F$_{\nu}$ $\propto$ ${\nu}^{-0.6}$, 
with a cooling frequency 3 $\times$ 10$^{11}$ GHz $<$ $\nu_{c}$ $<$ 
3 $\times$ 10$^{14}$ GHz.
Then we expect a slope between $\nu_{m}$ and $\nu_{c}$ of $\beta$ = 
(p$-$1)/2 = 0.68 $\pm$ 0.06, consistent with the observed value.
As we have already mentioned, if the $p$ = 2.3 jet model is correct, by 
this time (Dec 12.0 UT), the cooling break should be already below the 
optical band, with an optical synchrotron spectrum  $F_{\nu}$ $\propto$ 
$\nu^{-p/2}$  = $\nu^{-1.15}$ that is in agreement with our optical data 
($\beta = -1.05\pm 0.09$). Therefore our Dec 12.0 observations support a 
jet model with $p$ = 2.30 $\pm$ 0.07, marginally consistent with 
$p$ = 2.52 $\pm$ 0.13 as proposed by Galama et al. (2000a) on the basis
of a  fit to the multiwavelength spectra from the radio to the R-band data.

\section{Conclusion}

Most currently popular theories imply a direct correlation between star 
formation and GRB activity. How does GRB 991208 fit into this picture ?
The angular coincidence of the OA and the faint host argues against a
compact binary merger origin of this event (Fryer et al. 1999)  and in favor 
of the involvement of a massive star (Bodenhaimer and Woosley 1983, Woosley 
1993, Dar and De R\'ujula 2001). 
The very rapid photometric decline of the afterglow of GRB 991208 
provided hope for the detection of the much fainter light contamination 
from the underlying supernova, what we have confirmed on the basis of the 
late-time R-band measurements, thus giving further su\-pport to the 
massive star origin. 
There are still many unsolved riddles about GRBs, like the second break in
the light curve of this event, i.e. responsible for the steep decay seen 
after 5 days. 
The community continues
to chase GRB afterglows, and with every new event we make progress by
finding more clues and creating even more new puzzles.

\vspace{1cm}

{\it Acknowledgements.}
The Calar Alto German-Spanish Observatory is operated jointly by
the Max-Planck Institut f\"ur Astronomie in Heidelberg, and the Comisi\'on
Nacional de Astronom\'{\i}a, Madrid. The Sierra Nevada Telescope is operated 
by the Instituto de Astrof\'{\i}sica de Andaluc\'{\i}a (IAA). The Nordic 
Optical Telescope (NOT) is operated on the island of La Palma jointly by 
Denmark, Finland, Iceland, Norway and Sweden, in the Spanish Observatorio del 
Roque de los Muchachos (ORM) of the Instituto de Astrof\'{\i}sica de 
Canarias (IAC).  The data  presented here have been taken using ALFOSC,  
which is owned by the IAA and operated at the NOT under agreement between 
the IAA and the NBIfA of the Astronomical Observatory of Copenhagen.
This paper is also based on observations made with the Italian Telescopio 
Nazionale Galileo (TNG)
operated on the island of La Palma by the Centro Galileo Galilei of the CNAA
(Consorzio Nazionale per l'Astronomia e l'Astrofisica) at the Spanish
ORM of the IAC.
We thank P. Garnavich and A. Noriega-Crespo for making available to us 
the VATT image taken on Dec. 12.52 UT, A. Fruchter for his comments and
appreciate the generous allocation of observing time at the Calar Alto, 
Roque and Teide observatories. We are grateful to R. Gredel, U. Thiele, 
J. Aceituno, A. Aguirre, M. Alises, F. Hoyos, F. Prada and S. Pedraz for 
their support at Calar Alto, C. Packham (INT Group) for
his help to obtain the WHT spectra, the TNG staff for their support and
to J. M. Trigo (Univ. Jaime I) for pointing us the existence of the 
German meteor films.
KH is grateful for Ulysses support under JPL Contract
95805. J. Gorosabel acknowledges the receipt of a Marie Curie Research
Grant from the European Commission. This research was partially supported by 
the Danish Natural Science Research Council (SNF) and by a Spanish CICYT 
grant ESP95-0389-C02-02. 
V. V. Sokolov, T. A. Fatkhullin and V. N. Komarova  thank the 
RFBR N98-02-16542 ("Astronomy" Foundation grant 97/1.2.6.4) and INTAS 
N96-0315 for financial support of this work.

\end{document}